\documentstyle[11pt,twoside,jltp]{article}
\input epsf
\epsfverbosetrue

\title{Domain Wall Motion in the Presence of Nuclear Spins}

\author{M. Dub\'e$^{1}$ and P. C. E. Stamp$^{2,3}$}
\address{$^{1}$ Helsinki Institute of Physics, P.O. Box 9
(Siltavuorenpenger 20 C), Helsinki FIN-00014, Finland \\
$^{2}$ Department of Physics and Astronomy, and Canadian
Institute for Advanced Research,
University of British Columbia, Vancouver B.C. V6T 1Z1 Canada\\
$^{3}$ Institut Laue-Langevin, Ave. des Martyrs, Grenoble 38042, France \\
}

\runninghead{M. Dub\'e and P. C. E. Stamp}{Domain Wall Motion}
       
\begin{document}

\begin{abstract}
We investigate the motion of a domain wall 
in the presence of a dynamical hyperfine field. At temperature $T$ high 
compared to the hyperfine coupling, the nuclear spins create a
spatially random potential landscape, with dynamics dictated by the
nuclear relaxation time $T_2$. 
The distribution of the domain wall relaxation times (both in the
thermal and quantum regimes) can show a long tail, characteristic of
stochastic processes where rare events are important. Here, these
are due to occasional strong fluctuations in the nuclear
spin polarisation. 

PACS numbers: 75.45.+j
\end{abstract}

\maketitle


\section{Introduction}

The possibility that very large magnetic structures, such as domain walls 
and other magnetic solitons, might behave quantum mechanically, is of 
considerable current interest.
Many experiments in disordered magnets  
have given some indications
of domain wall tunneling, but these are difficult
to analyse (as is any other experiment 
in a disordered sample, either in the classical or quantum 
regimes). 
An obvious alternative is to perform experiments on samples 
containing only 1 or 2 domain walls, with a coercivity controlled by one or  
a few energy barriers. 
Since the tunneling process is itself stochastic, randomness is still present,
but this is of course inevitable in any quantum system.

However even in a completely pure system there is a source of {\it intrinsic}
randomness, coming from the nuclear spins, which couple strongly to any 
magnetic soliton.
In this paper, we examine some consequences of nuclear $T_2$
fluctuations for domain wall motion in general (including tunneling, already
discussed in \cite{dubejltp}). We do not propose a complete solution 
to the problem, but
we show how 
stochastic concepts, not generally associated with quantum
problems, may be relevant to the recent single wall tunneling experiments.

\section{Domain Wall Tunneling}

The motion of a smooth domain wall can in general be described by some
collective coordinate $q({\bf x}_{\bot},t)$ with ${\bf x}_{\bot}$
represent the coordinates in the plane of the wall. To this coordinate
is associated an effective mass $M_{w}$, arising from the
demagnetisation fields, and the force acting on the wall results from 
some potential $V(q)$. In many cases, due to the combined effects of
demagnetisation and surface energies, we can consider a flat wall,
described by a single coordinate $q(t)$, with an 
effective mass $M_w \sim S_w / (\gamma_{g}^{2} \lambda)$ where
$S_w$ is the surface of the wall, $\lambda$ its width, and 
$\gamma_g$ the gyromagnetic factor. 
A magnetic field $H_e$ adds a
linear ``pressure'' $\Delta V(q) \sim qH_e$, 
while a defect or non-magnetic impurity
yields an attractive potential $V(q) \sim
- U_o \mbox{sech}^2 (q/\lambda)$. 
The problem is then similar to the tunneling of a non-relativistic
particle in a potential $V(q)$, and 
the escape rate of a 
wall from the combined  defect-external field potential 
is given \cite{stampprl,IJMP92}, in the absence of
dissipation, by 
$\Gamma_0 \sim  \Omega_{0} \; \exp (-B_0 (\epsilon))$
where $B_0 (\epsilon) \sim N_{0} (H_c/M_0)^{1/2} \epsilon^{5/4}$ is the 
``bare '' WKB tunneling exponent, and $\Omega_{0} \sim 
(\gamma_{g}) (M_{0} H_{c})^{1/2} \epsilon^{1/4}$ is the small 
oscillation frequency of the wall in the potential. 
Here $N_0$ is the number of spins in the wall, $H_c$ the wall
coercive or ``escape'' field and $M_0$ the saturation magnetisation.
The control parameter $\epsilon = (1-H_e/H_c)$
goes to zero when $H_e = H_c$. Notice that $\Omega_0$ is also the
``bounce'' frequency of the problem, so that $\Omega_{0}^{-1}$ is the
barrier traversal time of the wall. 
Although these results were originally derived for the particular case
of a $180 ^{o}$ Bloch wall and short-ranged defect pinning potential, they 
have a more general applicability- for small $\epsilon$ almost 
any pinning potential acting on the wall 
will give similar results, and 
almost any magnetic
soliton
will have a similar tunneling exponent. 

Another quantity of experimental interest is the temperature $T_c$ at which
quantum tunneling starts to dominate over thermal activation. 
Usually $T_c \sim \Omega_0 / 2\pi$ but it is
important to realise that this only estimates a {\it crossover}
temperature (there can be no ``phase transition'') and that
any meaningful discussion of $T_c$ must include a
heat bath, and consequently, dissipation. For wall tunneling, detailed 
discussions have been given of the role of magnons \cite{stampprl,IJMP92},
electrons \cite{tata}, and phonons \cite{dubejltp}. A list of 
references, along with a brief review of both theory and experiments 
on domain wall tunneling up to mid-1997,
appears in the introduction of Dub\'e and Stamp \cite{dubejltp}.
In the present paper we wish to further develop
our preliminary discussion \cite{dubejltp} of the effect of 
the nuclear spin bath on 
magnetic wall dynamics.

Before doing so, we briefly consider the way in which the tunneling
experiments are done. We emphasize that the 
tunneling relaxation rate $\Gamma$  
it is not directly accessible in 
experiments- rather, one performs a series of ``trial'' tunneling experiments 
to establish
various probability distributions. 
The most common method is to ``ramp'' the applied magnetic 
field and record the 
field at which a tunneling event occurs. This yields the 
so-called ``switching field'' distribution, 
whose mean and the variance can be related to the
tunneling rate $\Gamma$. 
All the recent experiments on single domain wall tunneling have been 
based on this method. 

Let us consider 3 examples. 
Wernsdorfer et al. \cite{wernsdorfer} studied the process of 
{\it homogeneous} magnetisation reversal in single nickel wires (but
not the depinning process). 
By magnetoresistance measurements, Hong and
Giordano \cite{giordano} 
followed the depinning of a domain wall in a
nickel wire. The exact nature of the
soliton was not known, nor the nature of the pinning site
(most likely due to variations in the width of the wire, 
see ref. \cite{ono} for similar effects), 
but apparently quantum tunneling took 
place below $T_c \sim 2-3 \, K$. 
In a further set of experiments, they irradiated the sample
with microwaves, causing transitions between the energy levels of
the domain wall in the pinning potential, and increasing the
transition rate. 
Finally, Mangin et al. \cite{mangin} 
studied the propagation of a domain wall across
an energy barrier in a domain wall junction. 
Preliminary investigations
seems to indicate the possibility of quantum tunneling below 
$\sim 0.7 \, K$. For references to some other wall
tunneling experiments, see Dub\'e and Stamp \cite{dubejltp}.

\section{Nuclear Spins}

The set of nuclear spins ${\bf I}_k$ couples to the electronic spins 
${\bf s}_{k}$ by the hyperfine interaction, of strength 
$\omega_0 
\sim 1 \, mK \, - \, 0.5 \, K$ ). The {\it intrinsic} dynamics of
the nuclear spins comes from 
internuclear dipolar
interactions $V_{k k'}
\sim 1-100 \, \mbox{kHz}$,  
which cause flip-flop transitions
between the nuclear spins at a rate $T_2 \sim |V_{kk'}|^{-1}$.
The exchange energy between electronic spins transforms the microscopic
hyperfine interaction into an interaction between the domain wall and
all the nuclear spins in it. 
This total coupling 
can then be decomposed into 2 principal terms \cite{dubejltp}: 

{\bf (i)} A longitudinal potential $U(q)$ coming from the  
sum of the fields produced by the nuclear spins. Up to a time
$(\lambda/a_o)^2 T_2$ (where $a_o$ is the lattice parameter), 
the dynamic nuclear spin polarisation performs a
random walk, and the ensemble averaged correlation
${\cal C}_{UU}(q_{1}-q_{2}, t_1 -t_2)  =
\langle ( U(q_{1}(t_1))-U(q_{2}(t_2) )^{2} \rangle$) 
for a wall at 2 different positions and times is thus
\begin{equation}
{\cal C}_{UU}(q_{1}-q_{2}, t_1 -t_2) = \omega_{0}^{2} s^2 I^2 
{\cal E}^2 (q_1-q_2) \frac{|t_1 - t_2|}{T_2}
\label{correlations}
\end{equation}
where ${\cal E}^2 (q_1-q_2)$ is a function of the number of nuclear spins
swept by the domain wall between positions $q_1$ and $q_2$. 
At high temperatures,  $kT \gg \omega_{0}$, a volume containing $N$
nuclear spins has root mean square polarisation $\sim N^{1/2}$. 
One then finds that for a wall of surface $S_w$, 
in which a fraction $x$ of
all the states are occupied by nuclear spins,
${\cal E}^2 (q_{1}-q_{2}) \sim (x S_{w}/a_{0}^{3})(q_{1}-q_{2})$. 
However at very low T, when
$T \ll \omega_0$, the nuclear spins line up with the electronic spins in the 
wall, and 
${\cal E}(q_{1}-q_{2}) \sim x^2 S_{w}^{2} (q_{1}-q_{2})^{2} /a_{0}^{6}$.
Thus the wall is trapped in a potential which increases {\it linearly} (on
average) in both directions away from the wall centre. 

{\bf (ii)} There is also a transverse term, causing both
topological decoherence and dissipation in general \cite{stampjltp}. 
It describes the flipping of nuclear spins when the wall moves; one finds
that the number 
$\lambda_{I}$ of flipped spins caused by a single excursion between 2
points $q_{1}$ and $q_{2}$ (sweeping out $\Delta N$ nuclear spins) 
is $\lambda_{I} \sim (\Delta N/2)
( \pi \omega_0/2 \Omega_0)^2$, provided $\omega_{0} \ll \Omega_{0}$ (which 
will usually be the case in wall tunneling experiments).

\section{Tunneling in a Dynamically Fluctuating Potential}

The problem we are now interested in is the quantum dynamics of a 
particle in a 1-dimensional 
potential fluctuating randomly in time (in either the quantum 
or classical regimes). The problem of tunneling in such a potential seems 
hardly to have been considered before (we know of only 2 attempts 
\cite{dubejltp,bardou}). We consider here the specific example of  
a domain wall moving through the fluctuating nuclear field, where as we have 
just seen, the fluctuating nuclear component $U(q,t)$ 
adds on to a static or slowly
varying (ie., ramped) bare part $V(q)$. The nuclear 
component can either be weak
(as in materials like Ni and Fe), or much larger than the bare part 
(as in rare earth magnets).

In the {\it classical} regime, the decay rate simply depends on the height of
the potential barrier, 
which has fluctuations 
produced by the nuclear spins contained 
within the width (of order $\sim \lambda \epsilon^{1/2}$) of this barrier.
The total potential barrier height now has a Gaussian probability
distribution $P(V)$ 
with mean value $\tilde{V}$ and variance $E_0 = \omega_0 {\cal E}
(\lambda \epsilon^{1/2})$. 
This implies that the decay rate has a 
{\it log-normal} distribution \cite{lognormal}, ie., it is the quantity
$\ln (\Omega_0 /\Gamma)$ which is Gaussian distributed. Furthermore,
since $\Gamma$ cannot be larger than $\Omega_0$ by definition, this
distribution is truncated at a value $\xi = 1$ (the appearance of a broad
distribution has already been noted in a related problem \cite{bardou}, 
although the connection to the log-normal distribution was missed). 

The log-normal distribution has some very interesting properties.
If the ratio of the Gaussian variance to the Gaussian mean is
large enough, the
tail of the distribution is identical to a broad Levy distribution
\cite{levy} -
the stochastic process is dominated by rare events. 
Note however that in a genuine Levy distribution, the mean and/or variance 
are formally divergent, which is not the case with
the log-normal distribution- in the present problem the average thermal 
decay rate is 
\begin{equation}
\bar{\Gamma} = \Omega_0 \exp(-(\tilde{V}/T) + E_0^2/2 T^2)
\end{equation}

Although both mean and variance are well defined, 
the convergence of a series of measurements of $\Gamma$ (in a switching field 
experiment) to the mean value
of the distribution may be extremely slow, because of the truncation of the
distribution. This effect is most pronounced for large Gaussian 
variance \cite{lognormal,bardou}. 
Notice also that depending on the
desired degree of accuracy, a lognormal distribution may be 
approximated as a ``$1/f$'' distribution (ie., 
$P(\Gamma) \sim \Gamma^{-1}$), 
or to give an approximate ``stretched-exponential'' 
relaxation profile \cite{montroll}. 

In the {\it quantum} regime, the analysis of nuclear spin fluctuations is
similar, but slightly more technical \cite{dubejltp}. Since
the barrier traversal time $\Omega_0 \gg
T_2$, the tunneling process takes place in a
quasi-static nuclear spin potential. One can then introduce a 
``typical'' potential
$U_{\alpha} (q)$, where 
$U_{\alpha} (q) \sim \alpha E_0
(q/q_0)^{1/2}$
for $0 \le q \le q_0$, where the bare pinning potential $\tilde{V}(q)$ 
has a minimum at $q=0$, and a tunneling end-point is at $q=q_0$. 
The value $\alpha=1$ refers to the ensemble-averaged ``gaussian half-width''
value.
This simplification
allows the calculation of the tunneling in the limit
$E_0 \ll \tilde{V}$;
the tunneling exponent becomes 
$B(\epsilon,\alpha) \sim  B_0 (\epsilon) ( 1 +
(\alpha E_0/\tilde{V}(\epsilon))^{2/3} \, sign \, \alpha )$, the
appearance of the factor $\alpha^{2/3}$ coming from the shift of
the tunneling end-points by the nuclear potential. The distribution
of the decay rates is now narrower than the log-normal distribution, but the
qualitative behaviour is similar to the thermal case. 
The average decay rate is now $\bar{\Gamma} \sim \Gamma_0 (\epsilon)
\exp(+ \Delta \bar{B} (\epsilon))$ with
\begin{equation}
\Delta \bar{B} (\epsilon) \sim (E_0/\tilde{V} (\epsilon) ) B_{0}^{3/2}
(\epsilon)
\end{equation}
where $\Gamma_0 (\epsilon)$ is the bare tunneling rate. 
We emphasise that these results are valid for $E_0/\tilde{V}\ll 1$, a
condition not necessarily satisfied in rare-earth materials. They also ignore 
the dissipative effects of phonons, magnons, and electrons, which have been 
treated previously \cite{dubejltp,stampprl,IJMP92,tata}.

Let us now discuss the experimental consequences of these results. 
They apply directly to metals like Ni and Fe, where hyperfine 
interactions are weak. 
For all the experiments mentioned
above, it is  easy to verify that
the rate of change $d \tilde{V} /dt$ of the external ``ramped'' potential 
is always much smaller than the fluctuation rate
of the hyperfine potential. Since $T_2 << \bar{\Gamma}$, the
domain wall samples virtually  
the whole distribution of nuclear spin potentials
and the observed relaxation rate is then the 
$\bar{\Gamma}$ described above, in both quantum and classical regimes (although
in the classical regime there will be strong corrections coming from 
interactions with magnons \cite{stampprl,IJMP92}). 
We re-emphasize that the slow convergence
of the truncated log-normal distribution should be taken into account in 
analysing experiments. 

Finally, we also note that $T_1$ processes may also play a  
role at not too low $T$; moreover,
nuclear spin
fluctuations must certainly affect strongly the low-temperature Barkausen
noise 
spectrum in magnets (particularly rare-earth magnets); these points will be 
discussed elsewhere. 

\section{Acknowledgements}

M. D. thanks I. Koponen for very interesting discussions on L\'evy
distributions, and we both thank P. Nozieres for his hospitality in 
Grenoble.


\begin{thebibliography}{9}

\bibitem{dubejltp} M. Dub\'e and P. C. E. Stamp,
		   {\it J. Low Temp. Phys.} {\bf 110}, 779 (1998)

\bibitem{stampprl} P.C.E. Stamp, {\it Phys. Rev. Lett.} {\bf 66}, 2802 (1991).

\bibitem{IJMP92} P.C.E. Stamp, E.M. Chudnovsky, B. Barbara, 
{\it Int. J. Mod. Phys.}  {\bf B6}, 1355 (1992).

\bibitem{tata} G. Tatara and H. Fukuyama, {\it Phys. Rev. Lett.} 
{\bf 72}, 772 (1994); {\it J. Phys. Soc. Jap.} {\bf 63}, 2538 (1994).

\bibitem{wernsdorfer} W. Wernsdorfer et al, {\it Phys. Rev.} {\bf B55},
11552 (1997).

\bibitem{giordano} K. Hong and N. Giordano, {\it J. Phys. CM}  
{\bf 8}, L301 (1996), K. Hong, N. Giordano, 
{\it Europhys. Lett. 36}, 147 (1996).

\bibitem{ono} T. Ono et al., {\it Appl. Phys. Lett.} {\bf 72}, 1116
(1998).

\bibitem{mangin} S. Mangin et al., {\it Europhys. Lett.} {\bf 39}, 675 (1997).

\bibitem{stampjltp} N. V. Prokof'ev and P. C. E. Stamp, {\it J. Low Temp.
Phys.} {\bf 104}, 143 (1996).

\bibitem{bardou} F. Bardou, {Europhys. Lett.} {\bf 39}, 239 (1997); for
periodically oscillating potential, see eg., M. P. A. Fisher, 
{\it Phys. Rev.} {\bf B 37}, 75 (1988). 

\bibitem{lognormal} J. Aitchison and J. A. C. Brown, {\it The Lognormal
Distribution}, Cambridge University Press (1957).

\bibitem{levy} See eg., {\it Levy Flights and Related Topics in
Physics}, ed. M. F. Shlesinger, G. M. Zaslavsky and U. Frisch, Berlin
Springer (1995).

\bibitem{montroll} E. W. Montroll and J. T. Bendler, {\it J. Stat. Phys.}
{\bf 34}, 129 (1984)

\end{thebibliography}
\end{document}